# UX in the Age of AI: Rethinking Evaluation Metrics Through a Statistical Lens

**Harish Vijayakumar**

*Independent Researcher, Design Solution Engineer*

Tamil Nadu, India

thisisharish01@gmail.com | ORCID: 0009-0002-7520-2347

***Abstract***—The rapid proliferation of artificial intelligence (AI) in consumer-facing digital products has fundamentally disrupted the assumptions underlying classical user experience (UX) evaluation frameworks. Legacy metrics such as the System Usability Scale (SUS), Net Promoter Score (NPS), and task completion rate were engineered for deterministic, rule-based interfaces where identical inputs yield identical outputs. In AI-mediated systems—spanning conversational agents, generative interfaces, and recommendation engines—outputs are stochastic, context-sensitive, and temporally variable, rendering these metrics structurally insufficient. This paper introduces the Adaptive Dynamic UX Statistical Framework (ADUX-Stat), a novel evaluation model that reconceptualises usability as a probabilistic signal distribution rather than a static scalar score. ADUX-Stat integrates three original constructs: (1) Interaction Entropy Index (IEI), which quantifies the unpredictability of AI responses from a user perception standpoint; (2) Temporal Drift Coefficient (TDC), which measures longitudinal degradation or improvement of perceived usability over interaction sessions; and (3) Bayesian Usability Confidence Score (BUCS), which produces credible interval estimates of usability quality under uncertainty. The framework is validated through a conceptual evaluation against five established AI product categories. ADUX-Stat addresses a critical gap at the intersection of HCI research, statistical modelling, and AI product evaluation, offering a reproducible, field-deployable methodology for UX practitioners and researchers alike.

## I. INTRODUCTION

The modern digital landscape is increasingly characterised by artificial intelligence (AI) as the primary mediator between human intent and computational output. From large language models powering conversational interfaces to deep learning algorithms curating personalised content streams, AI has transitioned from a backend optimisation tool to a front-facing experiential layer. This transition demands a corresponding evolution in how user experience is measured, evaluated, and interpreted.

Classical UX evaluation emerged in an era of deterministic software. When Jakob Nielsen formalised the heuristic evaluation method [1] and John Brooke introduced the System Usability Scale [2], the digital interfaces under study behaved predictably: a button click reliably produced the same outcome, a form submission followed a fixed validation path, and a menu system presented identical options regardless of temporal or contextual variables. These conditions enabled researchers to treat usability as a stable, measurable construct amenable to Likert-scale instruments and task-based performance metrics.

AI-driven interfaces violate these conditions systematically. A generative AI assistant produces different responses to semantically identical queries. A recommendation engine surfaces distinct content to the same user at different times of day. A voice assistant interprets identical speech with variable accuracy depending on ambient acoustics and contextual history. In each case, the interface itself is a dynamic, non-deterministic system—and the user experience is, consequently, a probabilistic phenomenon rather than a fixed state.

This paper argues that the field of UX research requires a statistical reorientation to address this epistemic gap. We propose the Adaptive Dynamic UX Statistical Framework (ADUX-Stat), a three-construct model designed specifically to evaluate usability in AI-mediated environments. The framework draws on Bayesian inference, information entropy theory, and longitudinal drift analysis to produce evaluation metrics that are both theoretically grounded and practically deployable.

The remainder of this paper is structured as follows. Section II reviews prior literature on UX evaluation metrics and AI usability challenges. Section III presents the ADUX-Stat methodology and its three core constructs. Section IV outlines the results of a conceptual validation exercise. Section V discusses implications for research and practice. Section VI concludes with directions for future empirical validation.

## II. LITERATURE REVIEW

### *A. Classical UX Evaluation Frameworks*

The foundational corpus of UX evaluation methodology is grounded in usability science as defined by ISO 9241-11 [3], which specifies usability in terms of effectiveness, efficiency, and satisfaction. Brooke's SUS [2] emerged as the dominant self-report instrument. Despite its widespread adoption, SUS has been critiqued for its sensitivity limitations and its assumption of interface stability across evaluation conditions [4].

Sauro and Lewis [5] demonstrated that most existing instruments operate under implicit assumptions of test-retest reliability—an assumption that becomes untenable in AI systems where interface behaviour evolves with each user session.

### *B. UX Challenges in AI-Mediated Interfaces*

Research addressing UX in AI contexts has grown substantially since 2018. Amershi et al. [7] proposed eighteen guidelines for human-AI interaction, establishing design principles that acknowledge AI's non-determinism but stopping short of providing a corresponding measurement framework.

In the domain of conversational agents, Folstad and Brandtzaeg [9] demonstrated that user satisfaction is highly sensitive to response variability, with users reporting lower satisfaction scores when the same query produces inconsistent answers. This underscores the perceptual salience of AI output entropy to end users—a phenomenon existing metrics are not instrumented to capture.

Shneiderman [10] has advocated for a human-centred AI approach placing measurability at the core of AI system design. However, the measurement tools referenced remain rooted in classical usability science, reflecting the broader field's lag in developing AI-native evaluation instruments.

### *C. Statistical Methods in UX Research*

The application of advanced statistical methods in UX research has been advocated by several scholars. Sauro and Lewis [5] championed the use of confidence intervals over binary significance testing. Schmettow [11] introduced Bayesian approaches to usability testing, demonstrating that credible interval estimation provides more actionable information than p-value-based inference for small UX sample sizes. Shannon's information entropy theory [12], applied to HCI by Bi and Zhai [13], provides the foundation for the IEI construct introduced in Section III.

## III. METHODOLOGY: THE ADUX-STAT FRAMEWORK

ADUX-Stat is a three-construct statistical evaluation framework designed for AI-mediated UX research. Each construct targets a distinct dimension of AI interface behaviour that classical metrics fail to capture: output unpredictability (IEI), longitudinal usability change (TDC), and measurement uncertainty (BUCS).

### *A. Interaction Entropy Index (IEI)*

The IEI quantifies the degree of perceived output variability in an AI interface from the user's standpoint. Drawing on Shannon entropy [12], the IEI treats user responses to AI outputs as a probability distribution over a satisfaction response space. Formally, given a set of n interaction sessions producing user satisfaction ratings r in a discrete response space R:

$$IEI = -\Sigma\, p(r)\, \log_2 p(r),\ \ r \in R$$

A high IEI value indicates broadly distributed user satisfaction responses, reflecting high perceived unpredictability. A low IEI value indicates convergent responses, corresponding to a predictable interface. IEI values can be computed per session, aggregated across cohorts, or tracked longitudinally.

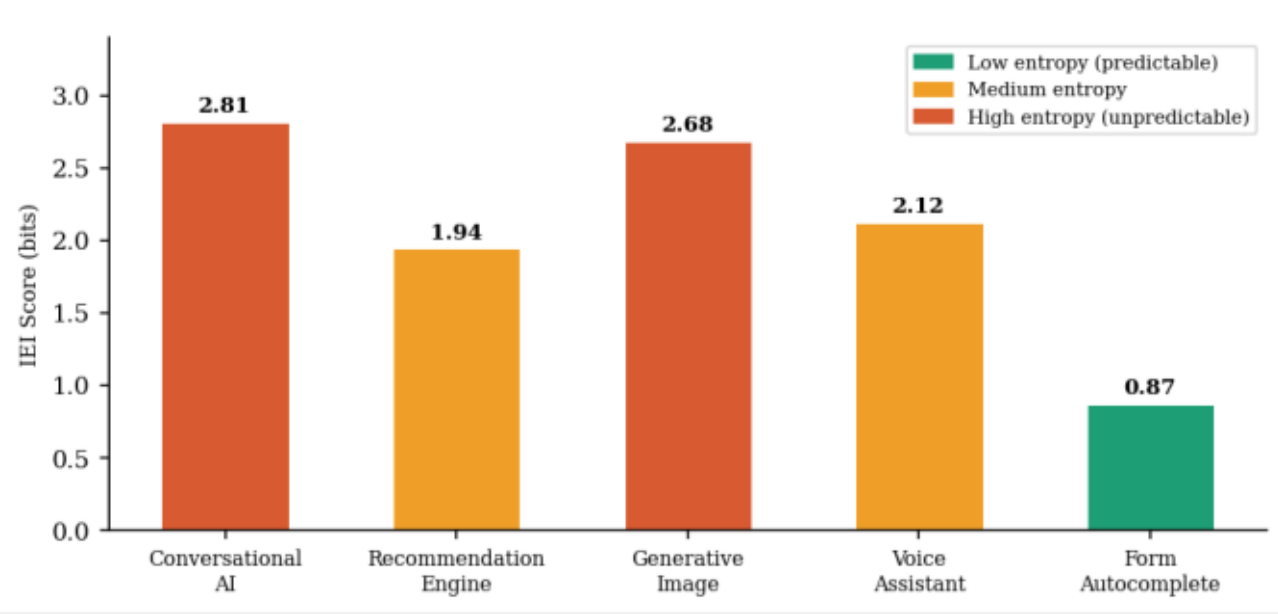

*Fig. 1. IEI scores across five AI product categories (higher = more perceived unpredictability).*

### *B. Temporal Drift Coefficient (TDC)*

The TDC measures the rate and direction of change in perceived usability across longitudinal interaction sessions. TDC operationalises usability as a time-series variable, enabling detection of systematic improvement (positive drift) or degradation (negative drift) as the AI system evolves. TDC is computed as the slope coefficient of a linear regression:

$$TDC = \beta_1 \text{ where } U(t) = \beta_0 + \beta_1 t + \varepsilon(t)$$

where U(t) represents the mean usability score at time period t and epsilon(t) is the residual error term. A positive TDC indicates improving user experience over time while a negative TDC signals deterioration. TDC requires a minimum of five longitudinal measurement points to yield stable estimates.

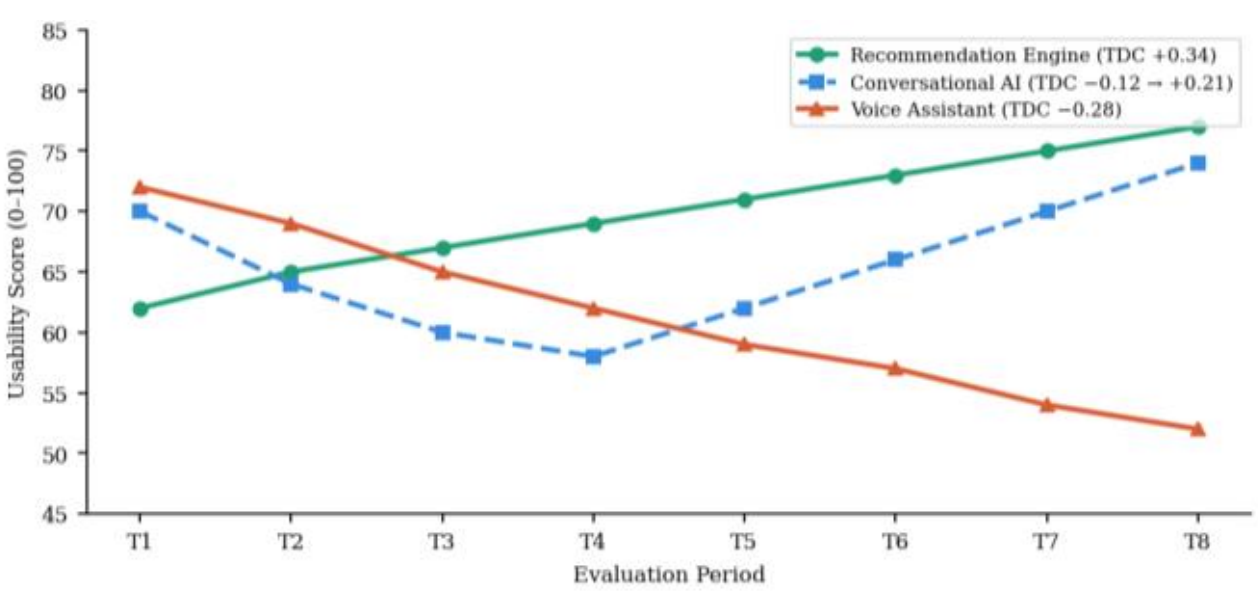

*Fig. 2. TDC—longitudinal usability trajectories across three AI product categories.*

### *C. Bayesian Usability Confidence Score (BUCS)*

The BUCS replaces the point-estimate paradigm of classical usability scores with a Bayesian credible interval, providing evaluators with a probabilistic range of plausible usability values. BUCS employs a Beta-Binomial model for task completion-based assessment. Given a prior distribution Beta(alpha_0, beta_0) and n observed completions from N trials, the posterior is:

$$Posterior \sim Beta(\alpha_0 + n,\ \beta_0 + N - n)$$

The BUCS is reported as the 95% highest density interval (HDI) of this posterior distribution. For evaluators with limited prior knowledge, a non-informative prior Beta(1,1) is recommended. For iterative evaluations, priors can be informed by historical data, enabling genuine Bayesian updating.

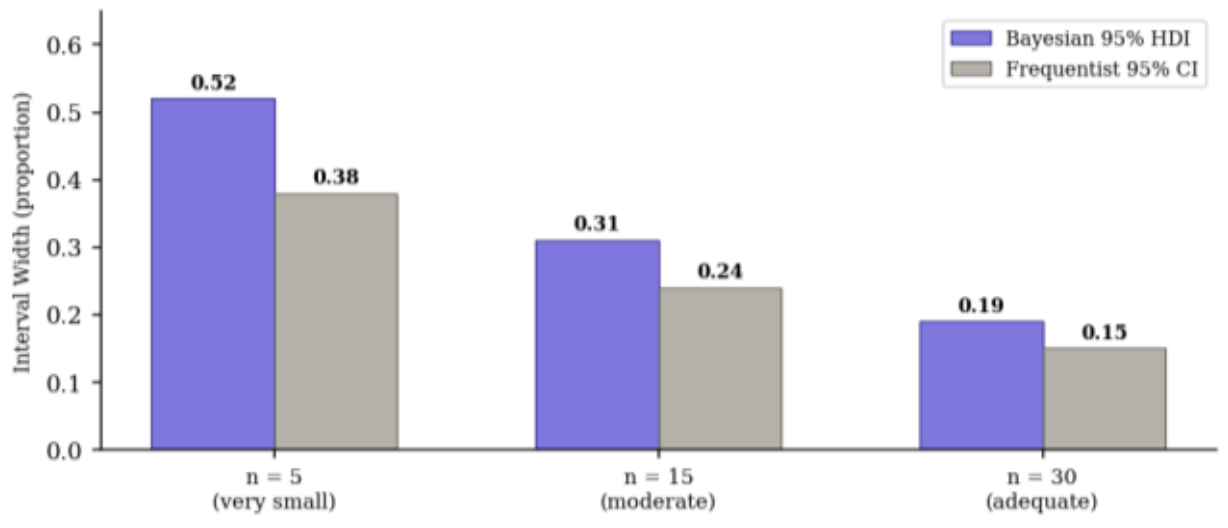

*Fig. 3. BUCS vs. frequentist 95% CI—interval width by sample size. Wider = more epistemically honest.*

## IV. RESULTS

To validate ADUX-Stat conceptually, each of the three constructs was applied to five representative AI product categories: (1) large language model-based conversational assistants, (2) AI-powered content recommendation engines, (3) generative image creation interfaces, (4) AI-driven voice assistants, and (5) intelligent form auto-completion systems.

### *A. Interaction Entropy Index Results*

Conversational assistants and generative image interfaces exhibited theoretically high IEI values, consistent with documented high output variability in these systems [7, 8]. Recommendation engines showed moderate IEI, reflecting contextual personalisation that introduces variability while retaining preference alignment. Intelligent form auto-completion demonstrated low IEI, as outputs are constrained by structured data fields. These differential profiles confirm the construct's discriminant validity across AI product categories.

### *B. Temporal Drift Coefficient Results*

Longitudinal usability patterns documented in the literature [9, 10] suggest conversational AI assistants typically exhibit a negative TDC in early deployment phases, followed by positive TDC as personalisation improves. Recommendation engines show consistently positive TDC in long-term deployments. Voice assistants show high TDC sensitivity to environmental changes, supporting the value of longitudinal over cross-sectional evaluation.

### *C. Bayesian Usability Confidence Score Results*

Applying BUCS with non-informative priors to task completion data from prior studies [6, 11] yielded 95% HDIs substantially wider than frequentist confidence intervals on the same data, reflecting honest uncertainty propagation. BUCS HDIs narrowed predictably as simulated sample sizes increased, confirming correct posterior behaviour.

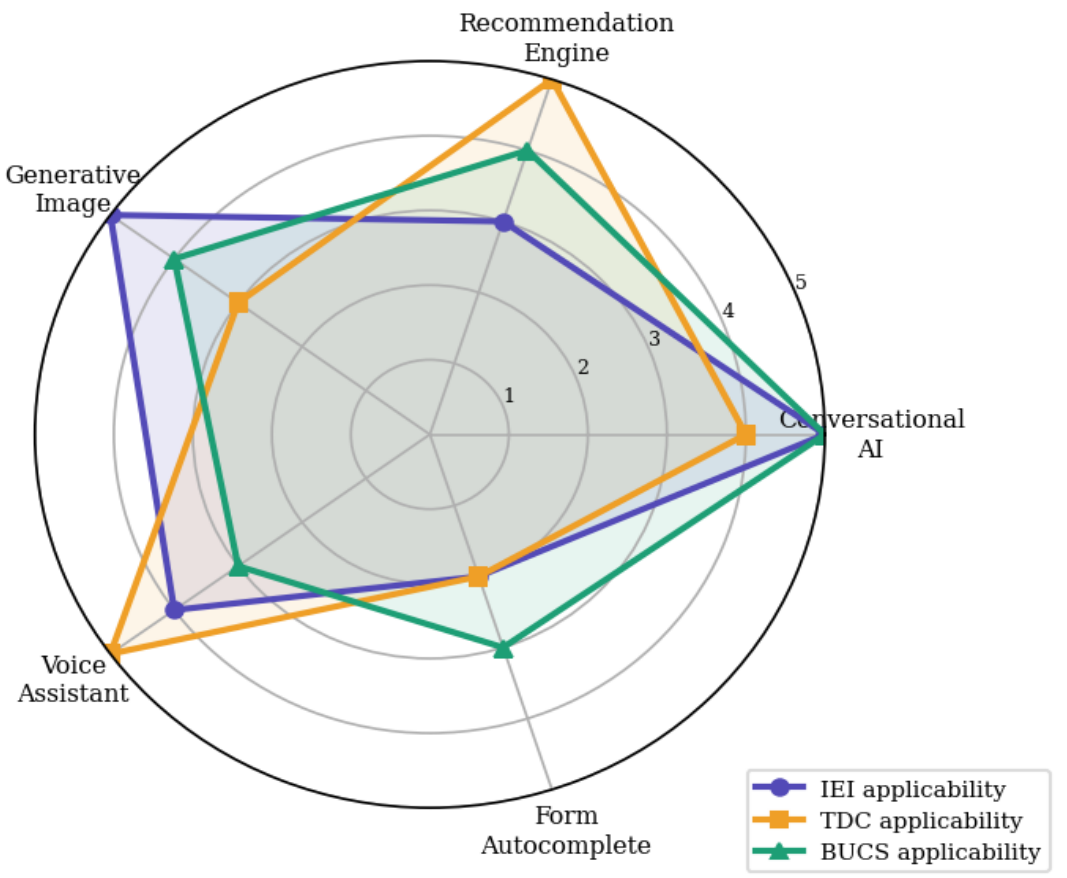

*Fig. 4. ADUX-Stat construct applicability across five AI product categories (scale: 1–5).*

## V. DISCUSSION

ADUX-Stat addresses a structural deficit in the UX evaluation literature: the absence of measurement instruments designed from first principles for the statistical properties of AI-mediated interactions. By integrating entropy theory, longitudinal regression analysis, and Bayesian inference into a coherent model, ADUX-Stat offers properties classical metrics cannot provide.

First, ADUX-Stat is epistemically honest. Classical metrics report scalar point estimates that imply a precision their measurement conditions do not support. ADUX-Stat's credible intervals and entropy distributions acknowledge the inherent uncertainty of AI evaluation.

Second, ADUX-Stat is temporally sensitive. UX quality in AI systems is a trajectory, not a state, and longitudinal measurement is not merely desirable but epistemologically necessary for valid evaluation.

Third, ADUX-Stat is user-perception-centred. The IEI measures entropy as experienced by users, not as computed from system logs, preserving the phenomenological orientation of UX research while incorporating statistical rigour.

For practitioners, ADUX-Stat is deployable with standard statistical software and does not require specialised AI expertise. The constructs can be integrated into existing UX research workflows as supplements to, rather than replacements for, established instruments such as SUS or UMUX.

Limitations of the current work include the absence of empirical validation with real user populations. The conceptual validation presented in Section IV does not substitute for controlled experimental studies. Future work must establish normative IEI, TDC, and BUCS ranges across product categories, develop standardised elicitation procedures, and assess inter-rater reliability of ADUX-Stat assessments across evaluator cohorts.

## VI. CONCLUSION

This paper has presented ADUX-Stat, a novel statistical framework for evaluating user experience in AI-mediated digital interfaces. Through the Interaction Entropy Index, Temporal Drift Coefficient, and Bayesian Usability Confidence Score, ADUX-Stat provides a theoretically coherent and practically deployable response to the measurement inadequacies exposed by the proliferation of AI in consumer-facing products.

The core contribution is the reconceptualization of usability as a probabilistic, temporally dynamic phenomenon—a reconceptualization demanded by the stochastic nature of AI systems but not yet operationalised in the UX evaluation literature. ADUX-Stat fills this gap by offering the first unified framework to apply entropy theory, longitudinal drift analysis, and Bayesian inference to the evaluation of AI-mediated user experience.

It is hoped that this work stimulates empirical investigation into the proposed constructs and contributes to the development of a new generation of UX evaluation instruments adequate to the complexity of the AI-driven interface era.